%% file: amc.tex
\documentclass[conference]{IEEEtran}
\IEEEoverridecommandlockouts

\usepackage[style=ieee,backend=bibtex]{biblatex}
\usepackage{amsmath,amssymb,amsfonts}
\DeclareMathOperator*{\argmax}{arg\,max}

\usepackage[utf8]{inputenc}
\usepackage[ruled,longend,linesnumbered]{algorithm2e}
\usepackage{textgreek}
\usepackage{amssymb}
\usepackage{graphicx}
\usepackage{textcomp}
\usepackage{xcolor}
\usepackage[acronyms]{glossaries}
\usepackage{svg}
\usepackage{mathtools}
\DeclarePairedDelimiter{\abs}{\lvert}{\rvert}
\DeclarePairedDelimiter\floor{\lfloor}{\rfloor}
\usepackage{breqn}   %
\usepackage{bm}
\usepackage[inline]{enumitem}
\usepackage{booktabs}
\usepackage{tabularx}
\input{acronyms}

\input{symbols}

\input{command}

\def\BibTeX{{\rm B\kern-.05em{\sc i\kern-.025em b}\kern-.08em
    T\kern-.1667em\lower.7ex\hbox{E}\kern-.125emX}}
\AtEveryBibitem{%
  \clearfield{note}%
  \clearfield{url}%
  \clearfield{urldate}%
}
\renewcommand\baselinestretch{.93}
\newcommand{\nonl}{\renewcommand{\nl}{\let\nl\oldnl}}
\let\oldnl\nl%
\begin{document}

\title{Adaptive Modulation and Coding based on Reinforcement Learning for 5G Networks
\thanks{This research was financed in part by the Coordenação de Aperfeiçoamento de Pessoal de Nível Superior - Brasil (CAPES) - Finance Code 001. \newline \indent This work was also supported by Ericsson Research, Technical Cooperation contract UFC.47. \newline \indent © 2019 IEEE.  Personal use of this material is permitted.  Permission from IEEE must be obtained for all other uses, in any current or future media, including reprinting/republishing this material for advertising or promotional purposes, creating new collective works, for resale or redistribution to servers or lists, or reuse of any copyrighted component of this work in other works.}
}

\author{\IEEEauthorblockN{Mateus P. Mota, Daniel C. Araújo, Francisco Hugo Costa Neto,\\ André L. F. de Almeida, F. Rodrigo P. Cavalcanti}
\IEEEauthorblockA{\textit{GTEL - Wireless Telecommunications Research Group} \\
\textit{Federal University of Ceará}\\
Fortaleza, Brazil \\
\{mateus, araujo, hugo, andre, rodrigo\}@gtel.ufc.br}
}
\maketitle

\begin{abstract}
We design a self-exploratory \gls{rl} framework, based on the Q-learning algorithm, that enables the \gls{bs} to choose a  suitable \gls{mcs} that maximizes the spectral efficiency while maintaining a low \gls{bler}.
In this framework, the BS chooses the \gls{mcs} based on the \gls{cqi} reported by the \gls{ue}.
A transmission is made with the chosen \gls{mcs} and the results of this transmission are converted by the \gls{bs} into rewards that the \gls{bs} uses to learn the suitable mapping from \gls{cqi} to \gls{mcs}.
Comparing with a conventional fixed look-up table and the outer loop link adaptation, the proposed framework achieves superior performance in terms of spectral efficiency and \gls{bler}.
\end{abstract}


\glsresetall

\section{Introduction}
\label{sec:Intro}

Link adaptation is a key enabling technology for broadband mobile internet, and has been part of the \gls{5g} \gls{nr} access technology.
In this context, \gls{amc} refers to the selection of the appropriate \gls{mcs} as a function of the channel quality, in order to keep the \gls{bler} below a predefined threshold.
In 4G long term evolution (LTE), the \gls{bler} target is fixed at 10\% \cite{3gpp.36.213}. However, 5G systems will cover a wider spectrum of services, requiring potentially different \gls{bler} targets \cite{Amin_2016,fantacci2009adaptive}.

\Gls{amc} is a good solution to match the link throughput to the time-varying nature of the wireless channel under mobility.
Periodically, the \gls{ue} measures the channel quality and maps this information into a \gls{cqi}.
The \gls{bs} uses the \gls{cqi} reported by the \gls{ue} to define the \gls{mcs}.
Typically, each \gls{cqi} is associated with a given \gls{snr} interval \cite{Blanquez-Casado2016}.
Considering \gls{lte} as an example, the \gls{bs} uses \gls{dci} embedded into the \gls{pdcch} to inform the \gls{ue} about each new \gls{mcs} selection \cite{ErikDahlman5G}.

Conventional solutions to the \gls{amc} problem includes the fixed look-up table \cite{fantacci2009adaptive}, also called \gls{illa}, and the \gls{olla} technique, which further improves the look-up table by adapting the \gls{snr} thresholds.
The \gls{olla} technique was first proposed in \cite{Sampath1997}, and was also addressed in \cite{Pedersen2007,Sarret2015, Blanquez-Casado2016}.

\Gls{ml} has become an attractive tool to devise novel \gls{amc} solutions in the context of complex emerging \gls{5g} systems and services.
In particular the drive towards self-organizing networks is potentially addressed by machine learning.
While in \gls{lte}, a look-up table provides fixed \gls{amc} rules for all the users, the emerging systems need a more flexible approach that can automatically adjust physical layer parameters (such as the modulation and coding scheme) according to the user channel state and service type.
\Gls{rl} refers to a category of \gls{ml} techniques \cite{survey-son} that has been applied to problems such as backhaul optimization~\cite{jaber2015}, coverage and capacity optimization~\cite{Fan2014} and resource optimization~\cite{Miozzo2017SwitchOnOffPF}.
There are few works that use \gls{rl} to solve the \gls{amc} problem.
In \cite{continuousState}, the selection of the \gls{mcs} is based on the received \gls{sinr}.
In this case, the state space is continuous, and the learning algorithm must handle a large state space.
In \cite{bruno2014robust} a Q-learning algorithm is proposed to solve the \gls{amc} problem in the context of a 4G \gls{lte} network.
A deep reinforcement learning approach is adopted in \cite{DRL_AMC} in the context of a cognitive heterogeneous network.

This work proposes a novel 5G \gls{amc} solution based on a \gls{rl} framework.
The proposed solution consists of collecting channel measurements at specific time instants to train an agent using the Q-learning algorithm.
The trained agent selects  a \gls{mcs} according to SNR measurements to maximize the current spectral efficiency.
We assume  a beam-based \gls{5g}-\gls{nr} as access technology, where the transmit and receive beams are selected using the beam sweeping procedure from \cite{giordani21}. The proposed \gls{amc} acts between any two consecutive points of sweeping.
We consider that the \gls{snr} between two consecutive points of sweeping tends to decrease due to the \ue~mobility  since it causes a mismatch among beams and the channel paths.
The agent uses the trained Q-table and the current measured \gls{snr} to properly select a \gls{mcs}.
To the best of authors' knowledge, previous works in \gls{amc} do not address the mismatch among beams and channel paths, while our solution works within the 5G-NR framework.

This work is structured as follows.
In Section \ref{sec:5gnr-trans} we briefly present the \gls{5g} \gls{nr} transmission model.
Section \ref{sec:system-model} describes the system and channel models used in this work.
In Section \ref{sec:proposed} we present the proposed \gls{amc} solution based on \gls{rl}.
Finally, Section \ref{sec:simulation} discusses our numerical results,
where the proposed \gls{rl} approach is compared against two baseline solutions, a fixed look-up table and an \gls{olla} algorithm.
The main conclusions are drawn in Section \ref{sec:conclusion}.

\section{Transmission Structure}
\label{sec:5gnr-trans}
\Gls{mac} uses services from the physical layer in the form of transport channels.
A transport channel defines the transmission over the radio interface, by determining its characteristics and how the information is transmitted \cite{3gpp.38.212} \cite{ErikDahlman5G}.
The transport channels defined for 5G-NR in the downlink are the \gls{dlsch}, \gls{pch}, and \gls{bch}.
In the uplink, only one transport-channel is defined, namely, the \gls{ulsch}.
Data transmissions in the downlink are carried out in the \gls{dlsch} and in the uplink the \gls{ulsch} \cite{AliZaidi632018}.
Data in the transport channel is organized into transport blocks.
At each \gls{tti}, up to two transport blocks of varying size are delivered to the physical layer and transmitted over the radio interface for each component carrier \cite{ErikDahlman5G}.

\gls{nr} supports \gls{qpsk} and three levels of quadrature amplitude modulation (16QAM, 64QAM and 256QAM), for both the uplink and downlink, with an additional option of $\pi/2$-BPSK in the uplink.
The \gls{fec} code in \gls{nr} for the \gls{embb} use case in data transmission is the \gls{ldpc} code, whereas in the control signaling polar codes are used.

The channel coding process in \gls{5g} \gls{nr} is composed of six steps \cite{ErikDahlman5G}, namely:
\gls{crc} attachment, \gls{cb} segmentation, per-\gls{cb} \gls{crc} attachment, \gls{ldpc} encoding, rate matching and \gls{cb} concatenation.

\section{System Model}
\label{sec:system-model}

\begin{figure}[tb]
\centerline{\includegraphics[width=\columnwidth]{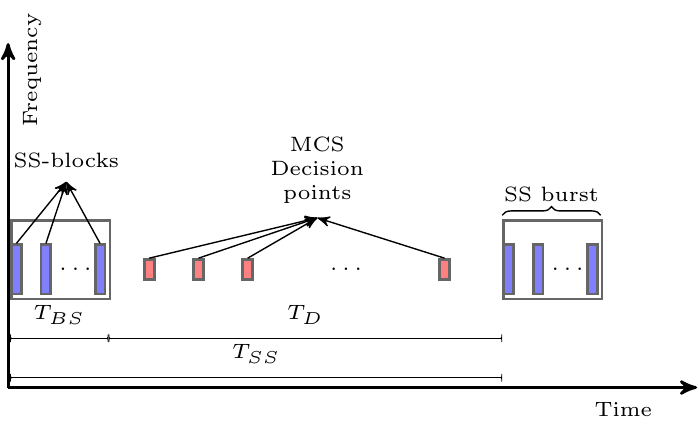}}
\caption{Model of time scheduling of operations.}
\label{fig:system-timing}
\end{figure}

Consider a single cell system whose \gls{bs} is equipped with \gls{not:txAnt} antennas serving one \gls{ue} with \gls{not:rxAnt} antennas. The signaling period, of duration $T_{SS}$ herein referred to as a \emph{frame}, is divided into two time windows, as shown in Figure \ref{fig:system-timing}. The first one contains a set of synchronization signal (SS) blocks with duration $T_{BS}$, where \emph{beam sweeping} is performed. More specifically, during this time window, the search for the best beam pair happens. The second time window is dedicated to data transmission using the selected beam pair. During this period, of duration $T_{D}$, the UE reports periodically the measured CQI to the BS that responds with the selected MCS.

During the transmission of the SS blocks, the BS measures all possible combinations of transmit and receive beams from the codebooks \gls{not:Wtx} \inSetComplex{\gls{not:txAnt}}{\gls{not:nBeams}} and \gls{not:Wrx} \inSetComplex{\gls{not:rxAnt}}{\gls{not:nBeams}}, respectively,  to select the beam pair with the highest \gls{snr}.
The selected beam pair for the $k$-th frame is expressed as
\begin{equation}
\label{eq.:beam_sweeping}
  \{ \bar{\mathbf{w}}_k, \bar{\mathbf{f}}_k \}= \argmax_{\mathbf{w}, \mathbf{f}} \frac{\|\mathbf{w}^H \mathbf{H}_t \mathbf{f}\|}{\sigma ^2},
\end{equation}
\noindent where $\mathbf{f}$ and $\mathbf{w}$ are columns of \gls{not:Wtx} and \gls{not:Wrx}, respectively,  $\channel _t $ \inSetComplex{\gls{not:rxAnt}}{\gls{not:txAnt}} is the channel between the \bs~ and the \ue at time $t$. We assume that the channel remains constant during the beam sweeping period $T_{BS}$.
The update of $\{ \bar{\mathbf{w}}_k, \bar{\mathbf{f}}_k \}$ depends on the periodicity $T_{SS}$ of the synchronization signal blocks, which can be  \{5, 10, 20 , 40, 80, 160\} (ms) \cite{giordani21}.
Therefore, the each beam pair solution remains constant within the time period $T_{SS}$, until the subsequent SS block arrives, when the BS can reevaluate Eq. \eqref{eq.:beam_sweeping}.

During the data transmission window, the discret-time received signal for the $t$-th symbol period associated with the $k$-th fixed beam pair, is given by
\begin{equation}
\label{eq.:rx_signal}
	\gls{not:Y}_{k,t} =
    \bar{\mathbf{w}}^H_k \,
  \channel _t\,
   \bar{\mathbf{f}}_k \,
   \gls{not:sscl}_t
 +
  \bar{\mathbf{w}}^H_k \;
  \gls{not:Z}_t,
\end{equation}

\noindent where $\gls{not:sscl}$ is the symbol transmitted to the \ue, and $\gls{not:Z}_t$ is the additive white Gaussian noise with zero mean and variance \gls{not:var}.
Defining
\begin{equation}
  \tilde{h}_{k,t} =     \bar{\mathbf{w}}^H_k \,
  \channel _t\,
   \bar{\mathbf{f}}_k \, ,
\end{equation}
as the effective channel at time $t$, associated with the chosen beam pair $\{ \bar{\mathbf{w}}_k, \bar{\mathbf{f}}_k \}$, the effective SNR at the \ue \, is given by
\begin{equation}
    \label{eq.:snr}
    \textrm{SNR} = \frac{ \abs{
     \tilde{h}_{k,t}
      }^2 }{\gls{not:var}} p_{\gls{not:sscl}},
\end{equation}
where $p_{\gls{not:sscl}}$ is the the power of transmitted symbol.

\subsection{Channel Model}

We assume a geometric channel model with limited number \gls{not:scatterers} of scatterers.
Each scatterer contributes with a single path between \gls{bs} and \gls{ue}. Therefore, the channel model can be expressed as
\begin{equation}\label{eq.:channelModel}
\channel _t= \sqrt{\pathloss}\sum_{i=0}^{\gls{not:scatterers} - 1 } \gls{not:beta}_i \strRx \anglePair{i,t}{ue} \hermitian{\strTx  \anglePair{i,t}{bs}} e^{ \mathrm{j} 2 \pi f_i tT_s} ,
\end{equation}
\noindent where $T_s$ is the \gls{ofdm} symbol period, $\pathloss$ denotes the pathloss, \gls{not:beta} is the complex gain of the $k$th path and $f_i$ is the Doppler frequency for the $i$th path.
The parameters \azm~$\in$ \range{0}{2\pi} and \elev~$\in$ \range{0}{\pi} denote the azimuth and elevation angles at the \bs \, (\gls{aod}) and the \ue \, (\gls{aoa}).
We assume a \gls{ura}, the response of which is written as:
\begin{equation*}
  \begin{split}
    \strTx \anglePair{i,t}{bs} = & \frac{1}{\sqrt{\gls{not:txAnt}}} \bigg[ 1, \expUraPhase{}{i,t}{bs}, \\ & \ldots ,\expUraPhase{(\gls{not:txAnt} -1 )}{i,t}{bs} \bigg],
  \end{split}
\end{equation*}
where \dist ~is the antenna element spacing, and \wave~is the signal wavelength. The array response at \gls{ue} can be written similarly.

The expression in \eqref{eq.:channelModel} can be expressed compactly as
\begin{equation}
\label{eq.:channelModelMtx}
\channel _t = \strRxMtx \diag{\gls{not:betaVec}_t} \hermitian{\strTxMtx},
\end{equation}
where $\gls{not:betaVec}_t = \left[ \gls{not:beta}_0 e^{ \mathrm{j} 2 \pi f_0 t T_s}, \ldots, \gls{not:beta}_{\gls{not:scatterers}-1} e^{ \mathrm{j} 2 \pi f_{S-1} t T_s} \right]$, and the matrices \strRxMtx~and \strTxMtx~are formed by the concatenation of array response vector at the \bs~and \ue, respectively.

\subsection{Transmission Model}
\label{subsec:trans}
\begin{figure}[tb]
\centerline{\includegraphics[width=40mm]{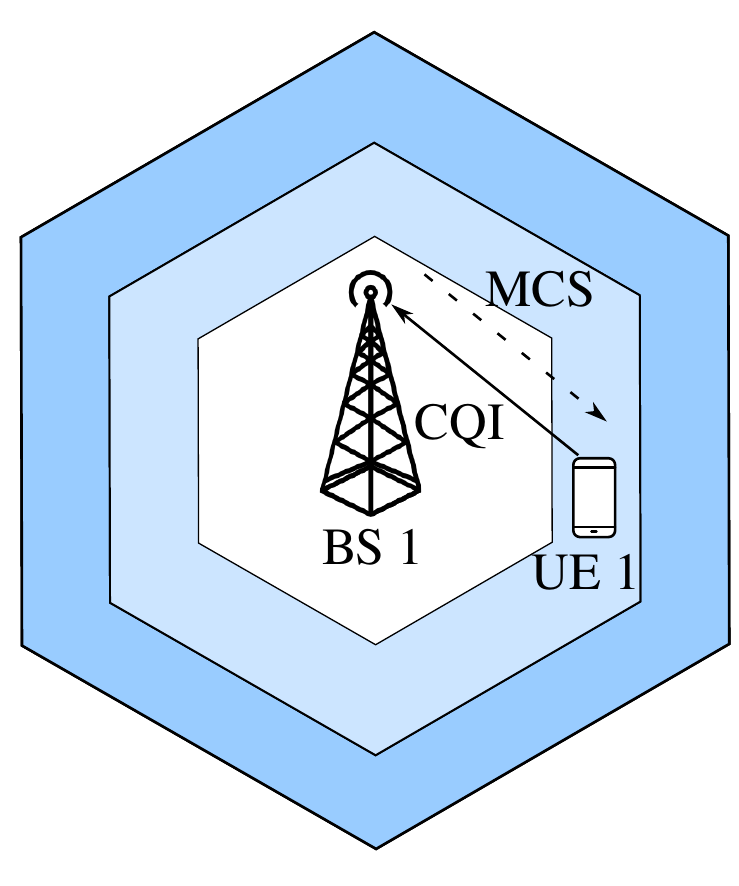}}
\caption{Exchange of signals involved in the AMC procedure}
\label{fig:system-model}
\end{figure}

The transmission process takes into account the channel coding and modulation blocks.
In this work, we implement all the steps specified in the \gls{nr} channel coding block except the rate matching \cite{3gpp.38.212}.
The \gls{cb} segmentation divides the transport block of $n_{bits}$ bits to fit the input size accepted by the \gls{ldpc} encoder, padding whenever necessary.
At the \gls{mcs} decision points, shown in Figure \ref{fig:system-timing}, the  \gls{ue} reports the measured \gls{cqi} to the \gls{bs}, which decides the \gls{mcs} accordingly.
The selected \gls{mcs} is informed to the \gls{ue} through the \gls{pdcch} as a part of the \gls{dci}. This process is shown in Figure \ref{fig:system-model}.

We considered a subset of the \gls{mcs}s in Table 5.1.3.1-1 in \cite{3gpp.38.214}, from the \gls{mcs} indexes 3 to 27. For our \gls{rl} based solution, the \gls{cqi} is a quantized measure of the \gls{snr}, and the number of possible \gls{cqi}s is defined by $N_{cqis}$. The \gls{cqi} metric for the \gls{rl}-\gls{amc} is defined as:
\begin{equation}\label{eq:cqi}
    CQI =
    \begin{cases}
    0, \text{if } SNR \leq SNR_{min}\\
    (N_{cqi}-1), \text{if } SNR \geq SNR_{max}\\
    \floor[\Big]{\frac{(SNR - SNR_{min})(N_{cqi}-1)}{SNR_{max}-SNR_{min}}}
    \end{cases}
\end{equation}
\noindent Note that each \gls{cqi}, except the minimum and the maximum ones, comprises \gls{snr} intervals having the same length.

At each \gls{tti} the \gls{bs} makes a transmission of a \gls{tb} of $n_{bits}$ at the chosen \gls{mcs}. The \gls{ue} receives a \gls{tb} from the \gls{bs} and, in possession of the chosen \gls{mcs}, decodes the \gls{tb} and calculates its \gls{ber}, \gls{bler} and spectral efficiency.
The \gls{bler} is the ratio of incorrectly received blocks over the total number of received blocks. %
The spectral efficiency $\eta$, in $bit/s/Hz$, is calculated as $(1-\gls{bler})\gls{not:mod}\gls{not:rate}$, where \gls{not:mod} is the number of bits per modulation symbol and \gls{not:rate} is the code rate.

\section{Q-Learning Based AMC}
\label{sec:proposed}

\subsection{Background on RL}

\Gls{rl} is a \gls{ml} technique that aims to find the best behavior in a given situation in order to maximize a notion of accumulated reward \cite{Bishop07}.
Unlike supervised learning, where the system learns from examples of optimal outputs, the \gls{rl} agent learns from trial and error, i.e., from its experience, by interacting with the environment.

\begin{figure}[tb]
\centerline{\includegraphics[width=80mm]{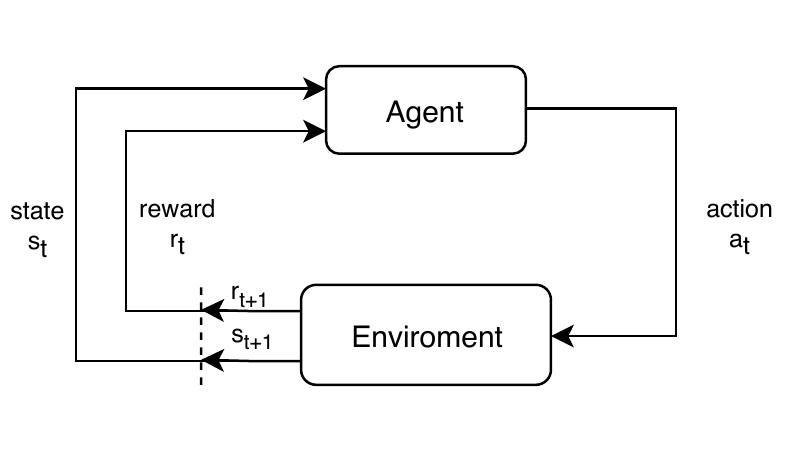}}
\caption{Basic diagram of a \gls{rl} scheme}
\label{fig:rlbasic}
\end{figure}

Figure \ref{fig:rlbasic} shows a simple block diagram of the \gls{rl} problem in which an agent, which is the learner and the decision maker, interacts with an environment by taking actions.
At each time step $t$, the agent receives the state $s_t$ of the environment and chooses an action $a_t$.
As consequence of its action, the agent receives a reward $r_{t+1} \in \mathcal{R} $, with $\mathcal{R} \subset \mathbb{R}$, and perceives a new state $s_{t+1}$ \cite{sutton2018rl}.
The goal of the \gls{rl} agent is to find the best policy that represents the best mapping of states to actions.
More specifically, the policy maps the perceived states of the environment to the action to be taken by the agent in those states.
The agent finds its best policy by taking into consideration the value of an action-value function.
The action-value function $Q^{\pi}(s_t,a_t)$, also known as Q-function, is the overall expected reward for taking an action $a_t$ in a state $s_t$ and then following a policy $\pi$.

One of the main paradigms in \gls{rl} is the balancing of \emph{exploration} and \emph{exploitation}.
There are different strategies to control the exploration- exploitation trade off. For a deeper discussion on this topic, we refer the interested reader to \cite{exploration2016}.
In this work, we make use of an adaptive $\epsilon$-greedy strategy, where the agent selects with probability $1-\epsilon$ the action with the higher action-value number, and with probability $\epsilon$ a random action. The $\epsilon$ parameter is initially set to a high value and is progressively decreased over time until a minimum value is reached.

In this work, we adopt the Q-learning algorithm \cite{Watkins1989}, which is an off-policy temporal difference (TD) algorithm \cite{sutton2018rl}.
The Q-learning algorithm works by updating its estimate of the action-value function based on each interaction of the agent with the environment.
The basic form of the action-values updates is given by Equation \eqref{QlearningEq}:

\begin{equation}\label{QlearningEq}
  \begin{split}
    Q\left(s_{t}, a_{t}\right) \leftarrow & (1-\alpha) Q\left(s_{t}, a_{t}\right) + \\
    & \alpha\left[r_{t+1}+\gamma \max _{a_{t+1} \in A} Q\left(s_{t+1}, a_{t+1}\right)\right],
  \end{split}
\end{equation}
\noindent where the parameter $0 \leq \alpha \leq 1$ is called \textit{learning rate} and the parameter $\gamma$ is called \textit{discount factor}, or discount rate, with $0 \leq \gamma \leq 1$.
The discount factor is used to control the importance given to future rewards in comparison with immediate rewards, so a reward received $k$ time steps later is worth only $\gamma^{k-1}$ times its value.

\subsection{Proposed AMC Solution}

The proposed solution is a Q-learning based link adaptation scheme, herein referred to as \gls{ql-amc}.
In the proposed approach, the \gls{bs} selects the \gls{mcs} based on the state-action mapping obtained from the Q-learning algorithm.
More specifically, the \gls{bs} chooses the \gls{mcs} using the Q-table obtained from the \gls{rl} algorithm.
The \gls{rl} based solution enables the system to learn the particularities of the environment and adapt to it.

\begin{figure}[t]
\centerline{\includegraphics[width=55mm]{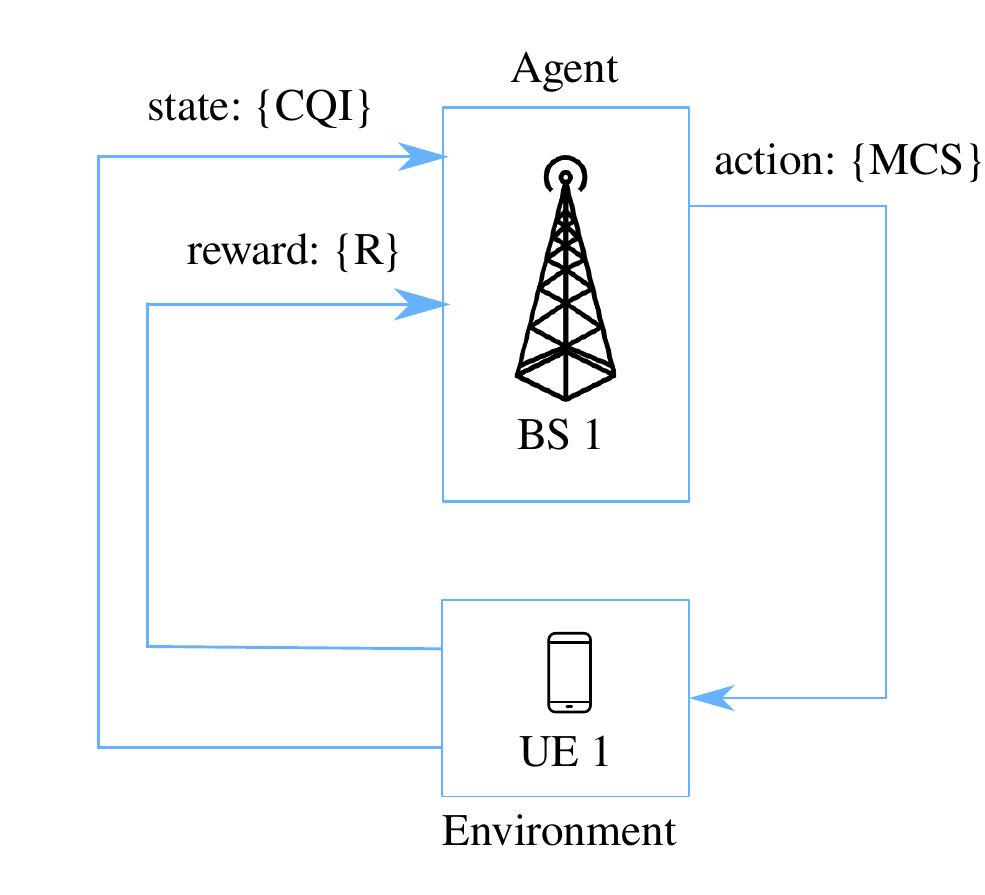}}
\caption{Basic diagram of the proposed AMC scheme}
\label{fig:rl-frame}
\end{figure}

A diagram adapting the model from Figure \ref{fig:rlbasic} to the \gls{amc} problem is shown in Figure \ref{fig:rl-frame}.

In the proposed \gls{amc} problem, the state space is the set of all possible \gls{cqi}s, from $0$ to $(N_{cqi}-1)$; the action space is the set of all possible \gls{mcs}s. As for the reward, we consider two different metrics.
The first reward function is a non-linear one defined as:
\begin{equation}\label{eq.:rewardBler}
  R_1 = \begin{cases}
  \gls{not:mod} \gls{not:rate}, \text{ if } BLER \leq BLER_T \\
  -1, \text{ else.}
\end{cases}
\end{equation}
\noindent where \gls{not:mod} is the number of bits per modulation symbol, \gls{not:rate} is the code rate and $BLER_T$ is the target \gls{bler} of the system, $10\%$ in case of \gls{embb} \cite{3gpp.38.214}.
The goal of this reward function is to allow the agent to choose the best \gls{mcs} that satisfies the \gls{bler} target. The second reward is defined in terms of the spectral efficiency (in bits/second/hertz):
\begin{equation}\label{eq.:rewardSE}
    R_2 = (1-BLER) \gls{not:mod} \gls{not:rate} \text{.}
\end{equation}
\noindent With this function, the agent will try to maximize the spectral efficiency.
A summary of the proposed \gls{ql-amc} algorithm is shown in Algorithm \ref{alg1}.

\section{Simulations and Results}
\label{sec:simulation}
\subsection{Simulation Parameters}
We assess the system performance with one \gls{bs} that serves one \gls{ue}.
The system has a bandwidth $B$ with a frequency carrier of $28$ GHz. Each resource block has a total of $12$ subcarriers and a subcarrier spacing $\gls{not:sub-spacing}= 120 \text{KHz}$.
We consider the channel model defined in \eqref{eq.:channelModel}.
The path loss follows a urban macro (UMa) model with non-line-of-sight (NLOS). Shadowing is modeled according to a log-normal distribution with standard deviation of $6$ dB \cite{AliZaidi632018}.
The noise power is fixed at $-123.185$ dBm.
A summary of the main simulation parameters is provided in Table~\ref{tab:sim-params}, while the parameters of the proposed \gls{ql-amc} algortihm are listed in Table \ref{tab:rl-params}.

\begin{table}[t]
\centering
\caption{Simulation Parameters}
\label{tab:sim-params}
\begin{tabularx}{0.8\columnwidth}{X r}
\toprule
\textbf{Parameter} 	& \textbf{Value} \\
\midrule
\gls{bs} height & 15 m\\
\gls{ue} height & 1.5 m\\
\gls{ue} track & rectilinear\\
\gls{bs}  antenna model & omnidirectional \\
\gls{bs}  antennas & 64 \\
\gls{ue} antenna model & omnidirectional \\
\gls{ue} antennas & 1 \\
Transmit power & 43 dBm\\
Frequency & 28 GHz\\
Bandwidth & 1440 MHz\\
Number of subcarriers  & 12\\
Subcarrier spacing & 120 kHz\\
Number of subframes & 10\\
Number of symbols & 14\\
Number of information bits per TTI & 1024\\
Azimuth angle spread & $[-60^{\circ}, 60^{\circ}]$\\
Azimuth angle mean & $0^{\circ}$\\
Elevation angle spread & $[60^{\circ}, 120^{\circ}]$\\
Elevation angle mean & $90^{\circ}$\\

Number of paths & 10\\
Path loss & UMa NLOS\\
Shadowing standard deviation & 6 dB\\
\bottomrule
\end{tabularx}
\end{table}

\begin{table}[tb]
	\centering
	\caption{QL-AMC Parameters}
	\label{tab:rl-params}
	\begin{tabularx}{0.8\columnwidth}{l r}
		\toprule
		\textbf{Parameter} 	   & \textbf{Value} \\
    \midrule
    $SNR_{min}$ for Eq. \eqref{eq:cqi} & $-5$ \\
    $SNR_{max}$ for Eq. \eqref{eq:cqi} & $40$ \\
		Discount factor ($\gamma$) & 0.10\\
		Learning rate ($\alpha$) & 0.90\\
		Maximum exploration rate ($\epsilon_{\max}$) & 0.50\\
    Minimum exploration rate ($\epsilon_{\min}$) & 0.05\\
    Cardinality of state space & $\{10,15,30,60 \}$\\
		\bottomrule
	\end{tabularx}
\end{table}

\begin{algorithm}[b]
  \caption{\strut QL-AMC}
    \label{alg1}
 \nonl Initialize $Q(s, a) = 0$, for all $s \in \mathcal{S}, a \in \mathcal{A}$\;
 \nonl \ForEach{MCS Decision Point (see Fig. \ref{fig:system-timing})}{
  The UE \emph{observes the state} $s: \gls{cqi}$\ and feeds it back to the BS;\\
  The BS \emph{takes an action} $a: MCS$ using the policy driven by $Q$ (e.g., $\epsilon$-greedy);\\
  The BS \emph{perceives a reward} $r$ (c.f. Eqs. (\ref{eq.:rewardBler}) or (\ref{eq.:rewardSE})) and observes the next state $s^{\prime}$\;
  The BS update the Q-table: $Q(s, a) \leftarrow (1-\alpha) Q(s, a) + \alpha [r + \gamma \max_a Q(s',a)]$\;
  $s \leftarrow s^{\prime}$\;
\nonl }

\end{algorithm}

\subsection{Baseline Solutions}

We compare the \gls{ql-amc} against the \gls{amc} based on a fixed look-up table \cite{fantacci2009adaptive} and also against the \gls{olla} technique from \cite{Pedersen2007}.
In the fixed look-up table approach, a static mapping of \gls{snr} to \gls{cqi} is obtained by analyzing the \gls{bler} curves and selecting the best \gls{mcs}, in terms of throughput, that satisfies the target \gls{bler} \cite{bruno2014robust}.
The process of analyzing the \gls{bler} curves gives the \gls{snr} thresholds that separate each \gls{cqi}, as such the \gls{snr} to \gls{cqi} mapping for the look-up table and the \gls{olla} algorithm is different from the \gls{ql-amc} defined in Eq. \eqref{eq:cqi}.
We assumed a direct mapping of \gls{cqi} to \gls{mcs}, i.e., each \gls{cqi} is mapped to one \gls{mcs} only .
The \gls{olla} technique consists of improveing the conventional MCS look-up table by adjusting the \gls{snr} thresholds according to the \gls{acknack} from previous transmissions.
This adjustment is made by adding an offset to the estimated \gls{snr} to correct the \gls{mcs}s.
The \gls{snr} that is transformed to \gls{cqi} is:
\begin{equation}
\textrm{SNR}_{olla} = \textrm{SNR} + \Delta_{olla}
\end{equation}
\noindent where the $\Delta_{olla}$ is updated at each time step according to the Eq. \eqref{eq.:olla} \cite{Blanquez-Casado2016}:
\begin{equation}\label{eq.:olla}
  \Delta_{olla} \leftarrow \Delta_{olla} + \Delta_{up} * e_{blk} - \Delta_{down} *(1 - e_{blk}),
\end{equation}
\noindent where $e_{blk} = 1$ in case of NACK, or $e_{blk} = 0$ if the transmission is successful.
The parameters $\Delta_{up}$, $\Delta_{down}$ and the target \gls{bler}, $BLER_{T}$, are inter-related. In fact, by fixing the $\Delta_{up}$ and the $BLER_{T}$, the $\Delta_{down}$ can be calculated as \cite{Pedersen2007}:
$$
\Delta_{down} = \frac{\Delta_{up}}{\frac{1}{BLER_{T}} -1}.
$$
The target \gls{bler} for the \gls{olla} algorithm is fixed at $0.1$, while we assume three values for $\Delta_{up}$: 0.01dB, 0.1dB and  1dB.

\subsection{Experiment Description and Results}

The experiment devised to assess the performance of the \gls{ql-amc} in comparison to the baseline solutions (look-up table and OLLA) is composed of two phases, namely the learning phase and the deployment phase.
We also evaluate the effect of the type of reward function considered (i.e., Eqs. \eqref{eq.:rewardBler} or \eqref{eq.:rewardSE}), and the different number of \gls{cqi}s. As such, each \gls{ql-amc} configuration is defined in terms of the cardinality of the state space and the reward function.
The action space is the set of all possible modulations orders and code rates, being the same for all configurations.

\subsubsection{Learning Phase}
In the first phase, the \gls{rl} agent populates the Q-table to learn the environment.
Each configuration of the \gls{ql-amc} passes through this phase only one time. Our simulation time starts with the \gls{ue} positioned at a radial distance of $20m$ from the \gls{bs}. The UE moves away from the \gls{bs} up to a distance of 100m. Then, the \gls{ue} comes back to its original position following the same path in the reverse direction.
The \gls{ue} has a speed of $5km/h$ and the simulation runs for a time equivalent to $160s$ of the network time, which corresponds to the transmission of 32.000 frames.

\subsubsection{Deployment phase}
The second phase uses the knowledge from the first phase, but with an $\epsilon$-greedy policy with a fixed value of $\epsilon = 0.05$, accordingly to the minimum value of the $\epsilon$-decreasing in the training phase.
The goal is to have an assessment of how the \gls{rl} agent performs in the long run.

In the deployment phase, we compare the proposed \gls{ql-amc} solution with the baseline solutions (look-up table and OLLA).
We perform $200$ Monte Carlo runs. At each run, the \gls{ue} starts at a random position between $25m$ and $90m$ of the \gls{bs}.
The \gls{ue} moves in a random rectilinear direction with a random speed between $10km/h$ and $20km/h$. This corresponds to a total of $K=125$ frames. Recall that each frame comprises a beam sweeping procedure, followed by data transmission jointly with a MCS selection procedure, as shown in Figure \ref{fig:system-timing}.

Table \ref{tab:deploy-results} summarizes the results in the deployment phase in terms of average values for each configuration of the \gls{ql-amc} and baseline solution.
The first column represents the type of solution adopted.
We consider three \gls{olla} schemes, denoted as \gls{olla} 1, 2 and 3, which consider $\Delta_{up}$ 0.01dB, 0.1dB and 1dB, respectively.
The conventional \gls{amc} with a fixed look-up table is denoted as "Table".
The second column represents the number of \gls{cqi}s and the type column represents the reward function used, defined by Eqs. \eqref{eq.:rewardBler}, \eqref{eq.:rewardSE}, and denoted as \gls{bler} and SE.

\begin{table}[tb]
\centering
\caption{Deployment Phase Results (Average over 200 runs)}
\label{tab:deploy-results}
\begin{tabularx}{\columnwidth}{l X X X X r}
  \toprule
  Type    & Cardinality &      Reward  &  BLER &     SE  &    BER \\
  \midrule
   QL-AMC &     10 &           BLER   & 0.0320 &  3.6700 & 0.0088 \\
   QL-AMC &     15 &           BLER   & 0.0306 &  3.3238 & 0.0087 \\
   QL-AMC &     30 &           BLER   & 0.0302 &  3.5594 & 0.0087 \\
   QL-AMC &     60 &           BLER   & 0.0306 &  3.8783 & 0.0087 \\
   QL-AMC &     10 &           SE     & 0.0306 &  3.9187 & 0.0086 \\
   QL-AMC &     15 &           SE     & 0.0301 &  3.8207 & 0.0085 \\
   QL-AMC &     30 &           SE     & 0.0310 &  3.9922 & 0.0086 \\
   QL-AMC &     60 &           SE     & 0.0311 &  4.1553 & 0.0086 \\
    Table &      - &           -      & 0.0311 &  3.8704 & 0.0088 \\
   OLLA 1 &      - &           -      & 0.0309 &  3.6700 & 0.0088 \\
   OLLA 2 &      - &           -      & 0.0330 &  1.8511 & 0.0090 \\
   OLLA 3 &      - &           -      & 0.0343 &  0.9999 & 0.0092 \\
  \bottomrule
\end{tabularx}
\end{table}

Analyzing Table \ref{tab:deploy-results}, we see that the two \gls{ql-amc} configurations presenting the best results in terms of spectral efficiency are those with cardinality 30 and 60, adopting the reward function $R_1$ of Eq. \eqref{eq.:rewardSE}.

\begin{figure}[tb]
\centerline{\includegraphics[width=\columnwidth]{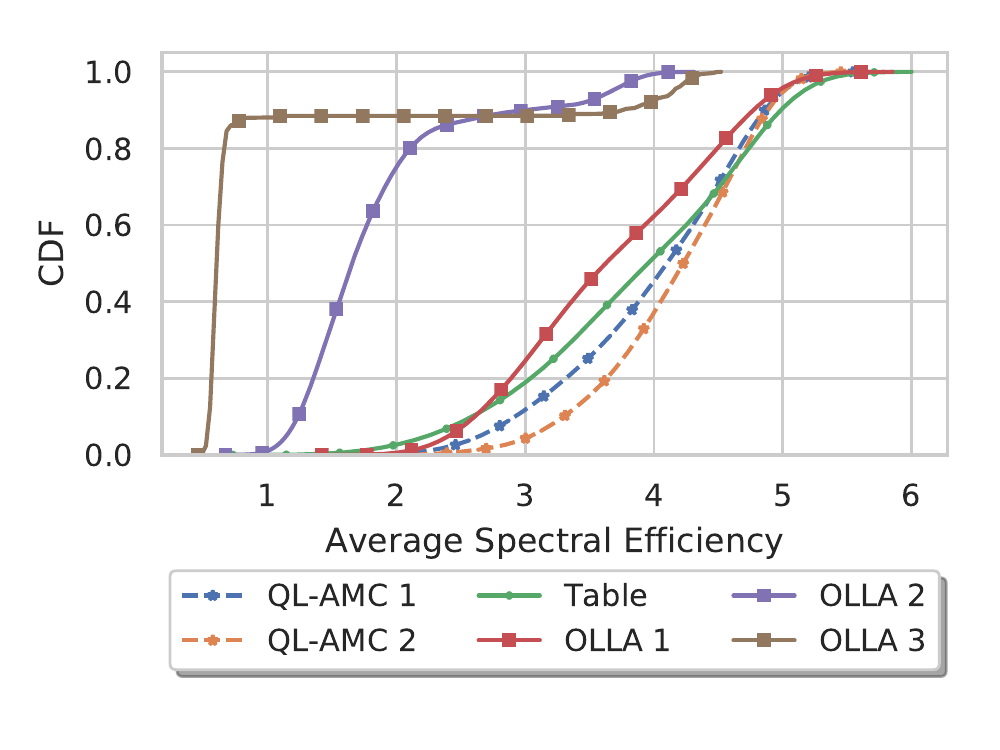}}
\vspace{-2ex}
\caption{CDF of average spectral efficiency (bps/Hertz)}
\label{fig:dep-spceff}
\end{figure}

Figure \ref{fig:dep-spceff} shows the cumulative distribution of the average spectral efficiency, in each Monte Carlo run, for the different \gls{ql-amc} configurations, with cardinality 30 and 60, which are labeled \gls{ql-amc} 1 and 2, respectively. We consider the reward function $R_2$ defined in Eq. \eqref{eq.:rewardSE}.
It can be seen that the proposed \gls{ql-amc} algorithm outperforms the baseline solutions in terms of spectral efficiency.

\section{Conclusions and Perspectives}
\label{sec:conclusion}
We demonstrate through simulations that the \gls{rl} provides a self-exploratory framework that enables the \bs~ to choose a suitable \gls{mcs} that maximizes the spectral efficiency.
Basically, the \gls{bs} decides a specific \gls{mcs} at a certain time instant. The \ue~ measures the reward of that action and report it to the \bs.
Comparing with the fixed look-up table and \gls{olla} solutions, the proposed QL-AMC solution has achieved higher spectral efficiencies and lower BLERs.
Between the two rewards considered, the second one that is in function of the spectral efficiency has achieved the best performance.
As a perspective, we highlight extensions to multi-layer MIMO transmission. Moreover, a comparison with other RL-based algorithms such as multi-armed bandits (MABs) \cite{zhou2015survey} or deep RL solutions \cite{DeepRLSurvey} is envisioned.

\renewcommand*{\bibfont}{\small}
\printbibliography

\end{document}

%% file: acronyms.tex
\newacronym{mmwave}{mmWave}{milimeter wave}
\newacronym{nr}{NR}{new radio}
\newacronym{5g}{5G}{fifth generation}
\newacronym{rl}{RL}{reinforcement learning}
\newacronym{bs}{BS}{base station}
\newacronym{ue}{UE}{user equipment}
\newacronym{3gpp}{3GPP}{Third Generation Partnership Project}
\newacronym{ss}{SS}{syncronization signal}
\newacronym{csi}{CSI}{channel state information}
\newacronym{rs}{RS}{reference signal}
\newacronym{ml}{ML}{machine learning}
\newacronym{mdp}{MDP}{markov decision process}
\newacronym{dlsch}{DL-SCH}{downlink shared channel}
\newacronym{bch}{BCH}{broadcast channel}
\newacronym{pch}{PCH}{paging channel}
\newacronym{ulsch}{UL-SCH}{uplink shared channel}
\newacronym{mac}{MAC}{medium acess control}
\newacronym{tti}{TTI}{transmission time interval}
\newacronym{crc}{CRC}{cyclic redundancy check}
\newacronym{qpsk}{QPSK}{quadrature phase shift keying}
\newacronym{qam}{QAM}{quadrature amplitude modulation}
\newacronym{ldpc}{LDPC}{low density parity check}
\newacronym{embb}{eMBB}{enhanced mobile broadband}
\newacronym{fec}{FEC}{forward error correction}
\newacronym{cb}{CB}{code-block}
\newacronym{aod}{AoD}{angles of departure}
\newacronym{aoa}{AoA}{angles of arrival}
\newacronym{ura}{URA}{uniform rectangular array}
\newacronym{sinr}{SINR}{signal-to-interference-plus-noise ratio}
\newacronym{snr}{SNR}{signal-to-noise ratio}
\newacronym{phy}{PHY}{Physical}
\newacronym{srs}{SRS}{sounding reference signals}
\newacronym{mu}{MU}{multi user}
\newacronym{mimo}{MIMO}{multiple-input multiple-output}
\newacronym{mcs}{MCS}{modulation and coding scheme}
\newacronym{cqi}{CQI}{channel quality indicator}
\newacronym{ber}{BER}{bit error rate}
\newacronym{bler}{BLER}{block error rate}
\newacronym{tb}{TB}{transport block}
\newacronym{amc}{AMC}{adaptive modulation and coding}
\newacronym{ql-amc}{QL-AMC}{Q-learning based adaptive modulation and coding}
\newacronym{illa}{ILLA}{inner loop link adaptation}
\newacronym{olla}{OLLA}{outer loop link adaptation}
\newacronym{dci}{DCI}{downlink control information}
\newacronym{pdcch}{PDCCH}{physical downlink control channel}
\newacronym{awgn}{AWGN}{additive white gaussian noise}
\newacronym{harq}{HARQ}{hybrid automatic repeat request}
\newacronym{lte}{LTE}{long term evolution}
\newacronym{acknack}{ACK or NACK}{positive or negative acknowledgments}
\newacronym{ofdm}{OFDM}{orthogonal frequency division multiplexing}

%% file: symbols.tex
\newglossary[nlg]{notation}{not}{ntn}{Notation}

      \newglossaryentry{not:nCell}{
  		name=\ensuremath{C},
  		description={number of cells},
  		type=notation}

    \newglossaryentry{not:rxAnt}{
    	name=\ensuremath{N},
        description={number of rx antennas},
        type=notation}

    \newglossaryentry{not:txAnt}{
    	name=\ensuremath{M},
        description={number of tx antennas},
        type=notation}

     \newglossaryentry{not:nUE}{
     	name=\ensuremath{L},
        description={number of users},
        type=notation}

      \newglossaryentry{not:l}{
     	name=\ensuremath{l},
        description={user index},
        type=notation}

    \newglossaryentry{not:nBeams}{
     	name=\ensuremath{K},
        description={number of beam pair per user},
        type=notation}

    \newglossaryentry{not:scatterers}{
        name=\ensuremath{S},
        description={number of scatterers},
        type=notation}

    \newglossaryentry{not:Wrx}{
     	name=\ensuremath{\mathbf{W}},
        description={codebook matrix at receiver side},
        type=notation}

    \newglossaryentry{not:Wtx}{
     	name=\ensuremath{\mathbf{F}},
        description={codebook matrix at transmitter side},
        type=notation}

    \newglossaryentry{not:c}{
	name=\ensuremath{c},
	description={cell index},
	type=notation}

    \newglossaryentry{not:i}{
     	name=\ensuremath{i},
        description={beam index at receiver side},
        type=notation}

    \newglossaryentry{not:I}{
        name=\ensuremath{I},
        description={number of beams of the rx codebook },
        type=notation}

    \newglossaryentry{not:j}{
     	name=\ensuremath{j},
        description={beam index at transmitter side},
        type=notation}

    \newglossaryentry{not:J}{
        name=\ensuremath{J},
        description={number of beams of the tx codebook },
        type=notation}

    \newglossaryentry{not:H}{
     	name=\ensuremath{\mathbf{H}},
        description={MIMO channel matrix},
        type=notation}

     \newglossaryentry{not:Herm}{
     	name=\ensuremath{H},
        description={Hermitian operator},
        type=notation}

     \newglossaryentry{not:y}{
     	name = \ensuremath{y_{\gls{not:i},\gls{not:j}}},
     	description={received measurement from rx beam \gls{not:i} with tx beam \gls{not:j}},
        type=notation}

    \newglossaryentry{not:Y}{
        name = \ensuremath{y},
        description={received measurement vector },
        type=notation}

    \newglossaryentry{not:z}{
        name = \ensuremath{z_{\gls{not:i},\gls{not:j}}},
        description={discrete Gaussian noise},
        type=notation}

    \newglossaryentry{not:Z}{
        name = \ensuremath{\mathbf{z}},
        description={discrete Gaussian noise vector},
        type=notation}

    \newglossaryentry{not:sscl}{
        name = \ensuremath{s},
        description={transmitted symbol},
        type=notation}

    \newglossaryentry{not:eye}{
        name = \ensuremath{\mathbf{I}},
        description={Identity matrix},
        type=notation}

    \newglossaryentry{not:var}{
        name = \ensuremath{\sigma ^2},
        description={Noise variance},
        type=notation}

    \newglossaryentry{not:beta}{
        name = \ensuremath{\beta},
        description={complex Gaussian random variable},
        type=notation}

    \newglossaryentry{not:betaVec}{
        name = \ensuremath{\bm{\beta}},
        description={vector complex Gaussian random variable},
        type=notation}

    \newglossaryentry{not:strRx}{
        name = \ensuremath{\mathbf{v}_{\textrm{\tiny{UE}}}},
        description={Rx sterring vector},
        type=notation
    }

     \newglossaryentry{not:strRxMtx}{
        name = \ensuremath{\mathbf{V}_{\textrm{\tiny{UE}}}},
        description={Set of Rx sterring vector},
        type=notation
    }

    \newglossaryentry{not:strTx}{
        name = \ensuremath{\mathbf{v}_{\textrm{\tiny{BS}}}},
        description={Tx sterring vector},
        type=notation
    }

     \newglossaryentry{not:strTxMtx}{
        name = \ensuremath{\mathbf{V}_{\textrm{\tiny{BS}}}},
        description={Set of Tx sterring vector},
        type=notation
    }

    \newglossaryentry{not:azm}{
        name = \ensuremath{\phi},
        description={Azimuth},
        type=notation
    }

    \newglossaryentry{not:elev}{
        name = \ensuremath{\theta},
        description={Elevation},
        type=notation
    }

    \newglossaryentry{not:pathLoss}{
        name = \ensuremath{\rho},
        description={Pathloss},
        type=notation
    }

    \newglossaryentry{not:wavelength}{
        name = \ensuremath{\lambda},
        description={Wavelength},
        type=notation
    }

    \newglossaryentry{not:dist}{
        name = \ensuremath{d},
        description={Distance between antenna elements},
        type=notation
    }

    \newglossaryentry{not:mod}{
        name=\ensuremath{\mu},
        description={modulation order},
        type=notation
    }

    \newglossaryentry{not:rate}{
        name=\ensuremath{\nu},
        description={code rate},
        type=notation
    }

    \newglossaryentry{not:sub-spacing}{
        name=\ensuremath{\Delta f},
        description={subcarrier spacing},
        type=notation
    }

%% file: command.tex
 \newcommand{\bs}{\gls{bs}}
 \newcommand{\ue}{\gls{ue}}

 \newcommand{\channel}{\gls{not:H}}
 \newcommand{\strRx}{\gls{not:strRx}}
 \newcommand{\strRxMtx}{\gls{not:strRxMtx}}
 \newcommand{\strTx}{\gls{not:strTx}}
 \newcommand{\strTxMtx}{\gls{not:strTxMtx}}
 \newcommand{\azm}{\gls{not:azm}}
 \newcommand{\elev}{\gls{not:elev}}
 \newcommand{\pathloss}{\gls{not:pathLoss}}

 \newcommand{\wave}{\gls{not:wavelength}}
 \newcommand{\dist}{\gls{not:dist}}

 \newcommand{\anglePair}[2]{\ensuremath{(\azm_{#1}^{#2}, \elev _{#1}^{#2} )}}

 \newcommand{\inSetComplex}[2]{\ensuremath{\in} \ensuremath{\mathbb{C}^{#1 \times #2} }}
 \newcommand{\hermitian}[1]{\ensuremath{#1 ^{H}}}
 \newcommand{\range}[2]{\ensuremath{\left[#1,#2\right]}}
 \newcommand{\expLetter}[1]{e^{#1}}
 \newcommand{\expUraPhase}[3]{\ensuremath{\expLetter{\left(\jmath #1 \frac{2\pi \dist}{\wave}(\sin{\azm _{#2}^{#3} }\sin{\elev _{#2} ^{#3}}+\cos{\elev _{#2} ^{#3}}) \right)}}}

 \newcommand{\diag}[1]{\textrm{diag}\left(#1\right)}